\documentclass[reprint,superscriptaddress,showpacs,nofootinbib,amsmath,amssymb]{revtex4-1}
\usepackage{comment}
\usepackage{amssymb}
\usepackage{color}
\usepackage{rotating}
\usepackage{amsmath}
\usepackage{ulem}
\usepackage{overpic}
\usepackage{graphicx}
\usepackage{dcolumn}
\usepackage{bm}
\usepackage{chngpage}
\usepackage{multirow}
\usepackage{slashed}
\usepackage{indentfirst}
\usepackage{booktabs}
\usepackage{amssymb,amsfonts}
\usepackage{amsmath}
\usepackage{color}
\usepackage[colorlinks,citecolor=blue,linkcolor=blue,hypertex,breaklinks=true]{hyperref}

\begin{document}
\title{Correlation between the charge radii difference in mirror partner nuclei and the symmetry energy slope}
\author{Xiao-Rong Ma}
\affiliation{School of Physics, Ningxia University, Yinchuan 750021, China}

\author{Shuai Sun}
\affiliation{Key Laboratory of Beam Technology of Ministry of Education, College of Nuclear Science and Technology, Beijing Normal University, Beijing 100875, China}

\author{Rong An}
\email[Corresponding author:]{rongan@nxu.edu.cn}
\affiliation{School of Physics, Ningxia University, Yinchuan 750021, China}
\affiliation{Key Laboratory of Beam Technology of Ministry of Education, College of Nuclear Science and Technology, Beijing Normal University, Beijing 100875, China}

\author{Li-Gang Cao}
\email[Corresponding author:]{caolg@bnu.edu.cn}
\affiliation{Key Laboratory of Beam Technology of Ministry of Education, College of Nuclear Science and Technology, Beijing Normal University, Beijing 100875, China}
\affiliation{Key Laboratory of Beam Technology of Ministry of Education, Institute of Radiation Technology, Beijing Academy of Science and Technology, Beijing 100875, China}

\date{\today}

\begin{abstract}
A correlation between the charge radii difference of mirror partner nuclei $\Delta{R_{\mathrm{ch}}}$ and the slope parameter $L$ of symmetry energy has been built to ascertain the equation of state of isospin asymmetric nuclear matter.
In this work, the influences of pairing correlations and isoscalar compression modulus on the $\Delta{R_{\mathrm{ch}}}$ are systematically investigated based on the Skyrme energy density functional theory.
The calculated results suggest that the linear correlation between $\Delta{R_{\mathrm{ch}}}$ and $L$ is decreased by the surface pairing correlations.
The slope parameter deduced from the difference of charge radii of mirror-pair nuclei $^{32}$Ar-$^{32}$Si, $^{36}$Ca-$^{36}$S, $^{38}$Ca-$^{38}$Ar, and $^{54}$Ni-$^{54}$Fe falls into the range of $L=42.57$-$50.64$ MeV, that is, the rather soft equation of state of asymmetric nuclear matter.
Besides, the range of the slope parameter can also be influenced by the effective forces classified by various isoscalar incompressibility coefficients.
\end{abstract}


\maketitle
\section{INTRODUCTION}
Nuclear symmetry energy is generally employed to comprehend the underlying physical mechanism from terrestrial nuclei to dense astrophysical events~\cite{STEINER2005325,LI2008113,ROCAMAZA201896,PhysRevC.100.045801,PhysRevC.101.034303,XuJianFeng}.
So far, plenty of methods have been applied to constrain the nuclear symmetry energy, such as the extracted neutron skin thickness (NST)~\cite{OYAMATSU19983,PhysRevLett.85.5296,FURNSTAHL200285,Liu_2011,ZHANG2013234,PhysRevC.102.044316}, properties of giant resonances~\cite{CaoLiGang_2008,PhysRevC.92.034308,PhysRevC.92.064304,PhysRevC.94.044313,Cheng_2023}, and quantities of the heavy-ion collision~\cite{LYNCH2009427,DiToro_2010,PhysRevC.84.044620,XIE20131510,FENG2010140,yuhao}.
However, a unified density-dependence of symmetry energy is hardly determined owing to the uncertainties of the different theoretical models.
Thus, alternative observables are urgently required for constraining the equation of state (EoS) of asymmetric nuclear matter.
The difference in charge radii of mirror-pair nuclei has been proposed as a tentative probe to evaluate the slope parameter of symmetry energy $L$~\cite{PhysRevC.88.011301,PhysRevLett.119.122502}.
Therefore, the precisely measured charge radii differences in mirror partner nuclei $^{32}$Ar-$^{32}$Si, $^{36}$Ca-$^{36}$S, $^{38}$Ca-$^{38}$Ar, and $^{54}$Ni-$^{54}$Fe have been thereby used to validate the range of the slope parameter $L$~\cite{PhysRevResearch.2.022035,PhysRevLett.127.182503,Konig:2023rwe}.

Charge radii differences in mirror pair nuclei are strongly associated with the neutron-skin thickness of neutron-rich nuclei and with the slope parameter $L$~\cite{PhysRevC.88.011301,PhysRevLett.119.122502}. As shown in Ref.~\cite{PhysRevC.97.014314}, the calculations performed using relativistic energy density functionals (EDFs) further inspect the validity of this assumption. It points out that the difference in proton radii of mirror nuclei can be indeed employed to constrain the EoS of isospin asymmetry nuclear matter.
This may provide an alternative approach to extract content information about the neutron skin thickness~\cite{CPCFangdeqing}.
A recent study suggests that the mirror-difference in nuclear charge radii is proportional to the isospin asymmetry $I=(N-Z)/A$, where $N$ and $Z$ represent the neutron and proton number of a nucleus, respectively, and $A=N+Z$ is the corresponding mass number~\cite{PhysRevLett.130.032501}.
Here, the same scenario can also be observed between the difference in binding energy per nucleon of mirror nuclei and the Coulomb asymmetry.
Furthermore, combing the accurately detected charge radii differences in mirror partner nuclei, the upper limit range of the symmetry energy slope has been determined to be $L\approx100$ MeV~\cite{PhysRevC.108.015802}.

This method requires more accurate charge radii data, which should be obtained through different nuclear models.
With the improvements in the experimental technique, high-precision charge radii of exotic nuclei can be detected~\cite{YANG2023104005}.
Thus, more charge radii data of nuclei far away from the $\beta$-stability line can be compiled~\cite{ANGELI201369,LI2021101440}.
This provides a fundamental guideline for theoretical studies to access the charge radii of mirror partner nuclei.
A reliable description of nuclear charge radii is influenced by various mechanisms, such as shape deformation~\cite{LALAZISSIS199635,PhysRevLett.117.172502,RongAn:054101,An_2023}, pairing correlations~\cite{PhysRevC.95.064328,PhysRevC.102.024307,PhysRevC.105.014325}, and shell closure effects~\cite{PhysRevC.100.044310,PhysRevC.104.064313}.
As is well known, pairing correlations play an important role in describing the ground state properties of finite nuclei~\cite{MENG19983,Geng:2003pk,RongAn:114101} or the quasiparticle resonant states~\cite{PhysRevC.102.054312}.
In particular, it has been mentioned that the radius of neutron-rich nuclei can also be influenced by the pairing correlations~\cite{PhysRevC.95.014316}.

As demonstrated in Ref.~\cite{PhysRevC.105.L021301}, the pairing correlations should be taken into account in calculating the nuclear charge radii.
As presented in Ref.~\cite{PhysRevC.107.034319}, the density energy functionals consisting of Skyrme and covariant models are used to systematically investigate the correlations between the charge radii differences in mirror partner nuclei and slope parameter $L$.
The calculated results suggest that the linear correlation between the mirror difference in charge radii and slope parameter $L$ is further decreased when considering the pairing effects.
In contrast, the correlation between the neutron-skin thickness of a neutron-rich nucleus and the proton radii difference of the corresponding mirror nuclei is mostly enhanced by pairing effects~\cite{CPCFangdeqing}.
Meanwhile, it has also been mentioned that the compression modulus of symmetric nuclear matter has an influence on determining the slope parameter $L$~\cite{An:2023toc}.

The density-dependence of nuclear symmetry energy, which is mostly determined by the slope parameter $L$, is associated with the isovector-sensitive indicators in the EoS of isospin asymmetry systems.
In Ref.~\cite{PhysRevC.69.024318}, it is stated that the correlation between the isoscalar incompressibility $K$ and the NST is not clear.
Further research demonstrated that the correlations between $K$ and the isovector parameters are generally weaker in comparison
with the strong correlation between NST and symmetry energy coefficients~\cite{PhysRevC.73.044320,PhysRevC.104.054324}.
Actually, the nuclear matter properties recognizing the isoscalar and isovector counterparts are mutually correlated with the effective model parameters.
This means that the influence coming from the isoscalar sector is non-negligible in discussing the isovector properties~\cite{PhysRevC.96.065805,PhysRevC.107.015801}.
As suggested in Refs.~\cite{PhysRevC.69.041301,Chen_2012,PhysRevC.104.055804}, the compression modulus of symmetric nuclear matter is sensitive to the density dependence of symmetry energy.
Therefore, it is also essential to investigate the correlation between the difference in charge radii of mirror-partner nuclei and the slope parameter $L$ under various incompressibility coefficients $K$.
In this work, the influences of pairing correlations and isoscalar compression modulus on the charge radii differences of mirror partner nuclei $^{32}$Ar-$^{32}$Si, $^{36}$Ca-$^{36}$S, $^{38}$Ca-$^{38}$Ar, and $^{54}$Ni-$^{54}$Fe are investigated using the spherical Skyrme EDFs. Furthermore, the range of the slope parameter $L$ derived from the charge radii differences of mirror-pair nuclei is discussed.

The contents of this paper are organized as follows.
Sec.~\ref{sec1} summarizes the theoretical framework.
In Sec.~\ref{sec2}, the numerical results and discussion are provided.
Finally, a summary and outlook are presented in Sec.~\ref{sec3}.

\section{THEORETICAL FRAMEWORK}\label{sec1}
The sophisticated Skyrme-type EDF, which is expressed as an effective zero-range force between nucleons with density-dependent and momentum-dependent terms, has achieved remarkable success in describing various physical phenomena~\cite{RevModPhys.75.121,PhysRevC.78.064306,PhysRevC.87.051302,PhysRevC.93.014302,PhysRevC.95.014316,jiang2018,PhysRevC.99.014314,
PhysRevC.100.014612,PhysRevC.101.014604,jiang2020,PhysRevC.102.054312,PhysRevC.90.024317,PhysRevC.102.014312,PhysRevC.102.014312,
PhysRevC.103.034321,PhysRevC.105.034315,WU2022136886,Huoenbo}.
In general, the effective force sets are determined by calibrating the properties of the finite nuclei and symmetric nuclear matter at saturation density $\rho_{0}$($\approx0.16~\mathrm{fm}^{-3}$).
Notably, it has been mentioned that some interactions can be used to characterize the bulk properties of finite nuclei and infinite nuclear matter (see, e.g.,~\cite{PhysRevC.68.014316,PhysRevC.85.035201} for details).
In this work, the Skyrme-like effective interaction has been recalled as follows~\cite{KOHLER1976301,CHABANAT1997710,CHABANAT1998231}:
\begin{eqnarray}
V(\mathbf{r}_{1},\mathbf{r}_{2})&=&t_{0}(1+x_{0}\mathbf{P}_{\sigma})\delta(\mathbf{r})\nonumber\\
&&+\frac{1}{2}t_{1}(1+x_{1}\mathbf{P}_{\sigma})\left[\mathbf{P}'^{2}\delta(\mathbf{r})+\delta(\mathbf{r})\mathbf{P}^{2}\right]\nonumber\\
&&+t_{2}(1+x_{2}\mathbf{P}_{\sigma})\mathbf{P}'\cdot\delta(\mathbf{r})\mathbf{P}\nonumber\\
&&+\frac{1}{6}t_{3}(1+x_{3}\mathbf{P}_{\sigma})[\rho(\mathbf{R})]^{\alpha}\delta(\mathbf{r})\nonumber\\
&&+\mathrm{i}W_{0}\mathbf{\sigma}\cdot\left[\mathbf{P}'\times\delta(\mathbf{r})\mathbf{P}\right].
\end{eqnarray}
Here, $\mathbf{r}=\mathbf{r}_{1}-\mathbf{r}_{2}$ and $\mathbf{R}=(\mathbf{r}_{1}+\mathbf{r}_{2})/2$ are related to the positions of two nucleons $\mathbf{r}_{1}$ and $\mathbf{r}_{2}$; $\mathbf{P}=(\nabla_{1}-\nabla_{2})/2\mathrm{i}$ is the relative momentum operator and $\mathbf{P'}$ is its complex conjugate acting on the left; and $\mathbf{P_{\sigma}}=(1+\vec{\sigma}_{1}\cdot\vec{\sigma}_{2})/2$ is the spin exchange operator that controls the relative strength of the $S=0$ and $S=1$ channels for a given term in the two-body
interactions, with $\vec{\sigma}_{1(2)}$ being the Pauli matrices.
The last term represents the spin-orbit force where $\sigma=\vec{\sigma}_{1}+\vec{\sigma}_{2}$.
The quantities $\alpha$, $t_{i}$, and $x_{i}$ ($i=0$-3) represent the parameters of the Skyrme forces used in this work.

Solving the Schr\"{o}dinger-like equations by self-consistent iteration leads to the eigenenergies and wave functions of the constituent nucleons. The pairing correlations can be generally treated either by the BCS method or by the Bogoliubov transformation~\cite{PhysRevLett.77.3963,MENG19983,Geng:2003pk}.
In this work, the spherical Skyrme Hartree-Fock-Bogoliubov (HFB) approach is utilized for all the calculations~\cite{BENNACEUR200596}.
The density-dependent zero-range pairing force is employed as follows~\cite{PhysRevC.86.054313}:
\begin{eqnarray}
V_{\mathrm{pair}}(\mathbf{r}_{1},\mathbf{r}_{2})=V_{0}\left[1-\eta\left(\frac{\rho(\mathbf{r})}{\rho_{0}}\right)\right]\delta(\mathbf{r}_{1}-\mathbf{r}_{2}),
\end{eqnarray}
where $\rho(\mathbf{r})$ is the baryon density in the coordinate space and $\rho_{0}=0.16~\mathrm{fm}^{-3}$ represents the nuclear saturation density. The value of $\eta$ is taken as $0.0$, $0.5$, or $1.0$ for volume-, mixed-, or surface-type pairing interactions, respectively. The quantity $V_{0}$ is adjusted by calibrating the empirical pairing gaps with the three-point formula~\cite{PhysRevC.86.054313,RongAn:114101}, and the values of $V_{0}$ are 188.2, 370.2, and 509.6 MeV fm$^{3}$ for the corresponding volume-, mixed-, and surface-type pairing interactions, respectively.

After achieving the convergence of the total binding energy, the root-mean square (rms) radii of the neutron and proton matter, and
the neutron-skin thickness can be obtained naturally.
The quantity of nuclear charge radius ($R_{\mathrm{ch}}$) can be calculated as follows (in units of fm$^{2}$):
\begin{eqnarray}\label{cp}
R_{\mathrm{ch}}^{2}=\langle{r_{\mathrm{p}}^{2}}\rangle+0.64~\mathrm{fm^{2}}.
\end{eqnarray}
The first term represents the charge density distributions of point-like protons and the second term is due to the finite size of protons~\cite{Gambhir:1989mp}. For mirror pair nuclei, the difference in charge radii ($\Delta{R_{\mathrm{ch}}}$) can be obtained through the formula mentioned above.

\begin{table*}[htbp!]
\centering
\caption{Parameters of the forces used in this work, where the parameter $\alpha=1/6$. Saturation properties of the Skyrme parametrization sets used in this work, such as the incompressibility coefficients $K$ (MeV), slope parameter $L$ (MeV), and symmetry energy $E_{\mathrm{sym}}$ (MeV) at saturation density $\rho_{0}$ (fm$^{-3}$), are also listed.}\label{tab0}
\begin{tabular}{ccccccccccccc}
\hline
\hline
~~~$K$~~~&~~~Sets~~~&~~~~$t_{0}$~~~~~& ~~~~~$t_{1}$~~~~&~~~~$t_{2}$~~~~&~~~~$t_{3}$~~~~& ~~~~$x_{0}$~~~~& ~~~~$x_{1}$~~~~&~~~~$x_{2}$~~~~&~~~~$x_{3}$~~~~& ~~~~$W_{0}$~~~~& ~~~~$L$~~~~& ~~~~$E_{\mathrm{sym}}$~~~~ \\
 \hline
& s3028 &$-2461.903$& 472.082& $-530.239$ &13599.53&  1.6830& $-0.3349$ & $-1.0$  & 2.6602&  124.766 &$-11.2262$ &   28   \\
& s3030 &$-2477.013$& 475.604& $-533.657$ &13709.09&  1.1408& $-0.3365$ & $-1.0$  & 1.8302&  117.482 &~~22.8715   &  30 \\
& s3032 &$-2491.849$& 486.614& $-579.661$ &13842.35&  0.9962& $-0.2791$ & $-1.0$  & 1.5763&  128.028 &~~36.2246   &  32 \\
$K\approx230$ & s3034 &$-2503.455$& 489.112& $-591.648$ &13939.40&  0.7308& $-0.2622$ & $-1.0$  & 1.1527&  126.846 &~~56.1442 &  34 \\
& s3036 &$-2513.951$& 491.018& $-615.662$ &14045.31&  0.5457& $-0.2210$ & $-1.0$  & 0.8437&  128.528 &~~71.5428   &  36  \\
& s3038 &$-2524.594$& 494.060& $-635.510$ &14142.73&  0.3550& $-0.1916$ & $-1.0$  & 0.5315&  129.200 &~~87.6155   &  38 \\
& s3040 &$-2531.688$& 489.598& $-625.976$ &14203.48&  0.1249& $-0.1940$ & $-1.0$  & 0.1662&  122.045 &~106.0862  &  40  \\
\hline
& s4028 &$-2296.534$& 515.586& $-336.464$ &11786.33&  1.7714& $-0.8491$ & $-1.0$  & 3.1703&  118.803 &~~~3.9774    &  28 \\
& s4030 &$-2317.546$& 532.898& $-355.019$ &11907.07&  1.2924& $-0.8449$ & $-1.0$  & 2.3824&  118.263 &~~34.0735   &  30  \\
& s4032 &$-2321.087$& 520.235& $-491.504$ &12164.33&  1.2626& $-0.5755$ & $-1.0$  & 2.1916&  136.989 &~~34.4283   &  32  \\
$K\approx240$& s4034 &$-2329.794$& 510.932& $-423.708$ &12179.54&  0.8546& $-0.6759$ & $-1.0$  & 1.5447&  117.932 &~~62.5884 &  34  \\
& s4036 &$-2337.516$& 507.834& $-424.568$ &12253.22&  0.7156& $-0.6664$ & $-1.0$  & 1.2882&  118.414 &~~75.6679   &  36  \\
& s4038 &$-2355.765$& 523.051& $-465.145$ &12389.16&  0.3790& $-0.6244$ & $-1.0$  & 0.7299&  118.238 &~~98.6522   &  38  \\
& s4040 &$-2359.001$& 513.325& $-475.563$ &12466.46&  0.2959& $-0.5827$ & $-1.0$  & 0.5558&  119.406 &~108.1741  &  40   \\
\hline
& s5028 &$-2157.179$& 599.608& $-396.271$ &10352.38&  1.4017& $-0.9193$ & $-1.0$  & 2.9016&  138.874 &~~33.0037   &  28  \\
& s5030 &$-2155.199$& 570.378& $-299.172$ &10324.29&  1.6366& $-1.0334$ & $-1.0$  & 3.2711&  131.218 &~~30.0248   &  30  \\
& s5032 &$-2172.510$& 583.512& $-348.992$ &10472.92&  1.4649& $-0.9688$ & $-1.0$  & 2.9140&  141.589 &~~43.5871   &  32  \\
$K\approx250$& s5034 &$-2179.610$& 584.069& $-331.053$ &10501.79&  1.2593& $-0.9977$ & $-1.0$  & 2.5371&  141.094 &~~60.3202  &  34 \\
& s5036 &$-2195.402$& 599.442& $-365.556$ &10609.25&  0.9683& $-0.9629$ & $-1.0$  & 2.0020&  145.638 &~~80.1762   &  36  \\
& s5038 &$-2205.358$& 602.093& $-375.998$ &10690.19&  0.7284& $-0.9493$ & $-1.0$  & 1.5517&  145.124 &~~97.4925   &  38  \\
& s5040 &$-2214.249$& 600.714& $-404.383$ &10802.96&  0.5278& $-0.9019$ & $-1.0$  & 1.1569&  146.358 &~112.2079  &  40  \\
\hline\hline
\end{tabular}
\end{table*}
The density dependence of symmetry energy can be expanded around the saturation density $\rho_{0}$ as follows:
\begin{equation}\label{eq:1}
E_{\mathrm{sym}}(\rho)\approx{E_{\mathrm{sym}}}(\rho_{0})+\frac{L}{3}\left(\frac{\rho-\rho_{0}}{\rho_{0}}\right)+\frac{K_{\mathrm{sym}}}{18}\left(\frac{\rho-\rho_{0}}{\rho_{0}}\right)^{2}+\dots,
\end{equation}
where $L$ and $K_{\mathrm{sym}}$ are the slope and curvature of symmetry energy at nuclear saturation density $\rho_{0}$, respectively.
These two quantities are defined as
\begin{eqnarray}
L&=&3\rho_{0}\frac{\partial{E_{\mathrm{sym}}(\rho)}}{{{\partial\rho}}}\Large{|}_{\rho=\rho_{0}},\\
K_{\mathrm{sym}}&=&9\rho_{0}^{2}\frac{\partial^{2}E_{\mathrm{sym}}(\rho)}{{\partial\rho^{2}}}\Large{|}_{\rho=\rho_{0}}.
\end{eqnarray}
Further, the isoscalar incompressibility $K$ of symmetric nuclear matter is recalled as
\begin{eqnarray}
K=9\rho_{0}^{2}\frac{\partial^{2}E_{0}(\rho)}{{\partial\rho^{2}}}\Large{|}_{\rho=\rho_{0}}.
\end{eqnarray}
Here, the quantity $E_{0}(\rho)$ is the energy per particle of symmetric nuclear matter~\cite{CPCFangdeqing}.
The nuclear incompressibility can be extracted from the measurements of the isoscalar giant monopole resonance
(ISGMR) in medium-heavy nuclei~\cite{PhysRevC.70.024307,PhysRevC.86.054313}.
In addition, it is also mentioned that the incompressibility of symmetric nuclear matter can be deduced from
the $\alpha$-decay properties~\cite{PhysRevC.74.034302}.
Table~\ref{tab0}, presents the parameters of the Skyrme forces used in this work. The corresponding saturation properties in infinite nuclear matter derived from the Skyrme forces, such as isoscalar incompressibility $K$, slope parameter $L$, and symmetry energy $E_{\mathrm{sym}}$, are also listed explicitly.
Both the slope parameter $L$ and symmetry energy $E_{\mathrm{sym}}$ at saturation density cover a wide range.

\section{results and discussion}\label{sec2}
The linear correlation between the charge radii differences in the mirror partner nuclei $^{32}$Ar-$^{32}$Si, $^{36}$Ca-$^{36}$S, $^{38}$Ca-$^{38}$Ar, and $^{54}$Ni-$^{54}$Fe and the slope parameter of symmetry energy have been sequentially employed to constrain the EoS of isospin asymmetric nuclear matter~\cite{PhysRevResearch.2.022035,PhysRevLett.127.182503,Konig:2023rwe}.
\begin{table}[!htb]
\caption{$R_{\mathrm{ch}}$ and $\Delta{R_{\mathrm{ch}}}$ database for the $A = 32$, $36$, $38$, and $54$ mirror-pair nuclei. The parentheses on the values of charge radii and the difference of charge radii show the systematic uncertainties~\cite{ANGELI201369,Miller2019,PhysRevLett.127.182503,PhysRevResearch.2.022035,Konig:2023rwe}.}
\label{tab1}
\doublerulesep 0.1pt \tabcolsep 9pt
\begin{tabular}{cccc}
\hline
~~~~$A$~~~~ &~~~~~ & ~~~~$R_{\mathrm{ch}}$~(fm)~~~~ & ~~~~$\Delta{R_{\mathrm{ch}}}$~(fm)~~~~ \\
\hline
32  & Ar &  3.3468(62) &  \\
    & Si  &  3.153(12)~~ & 0.194(14)\\
36  & Ca &  3.4484(27) &  \\
    & S  &  3.2982(12) & 0.150(4)~\\
38  & Ca &  3.4652(17) & \\
    & Ar &  3.4022(15) & 0.063(3)~ \\
54  & Ni &  3.7370(30) & \\
    & Fe &  3.6880(17) & 0.049(4)~ \\
\hline
\end{tabular}
\end{table}
In Table~\ref{tab1}, the experimental data of $R_{\mathrm{ch}}$ and $\Delta{R_{\mathrm{ch}}}$ for $A = 32$, $36$, $38$ and $54$ mirror-pair nuclei are given.
As mentioned above, the nuclear charge radius is influenced by the pairing correlations~\cite{PhysRevC.95.064328,PhysRevC.102.024307,PhysRevC.105.014325}.
Thus, the influence on the correlations between the charge radii difference in mirror partner nuclei and the slope parameter $L$ should be systematically investigated.

\begin{figure}[htbp]
\includegraphics[scale=0.33]{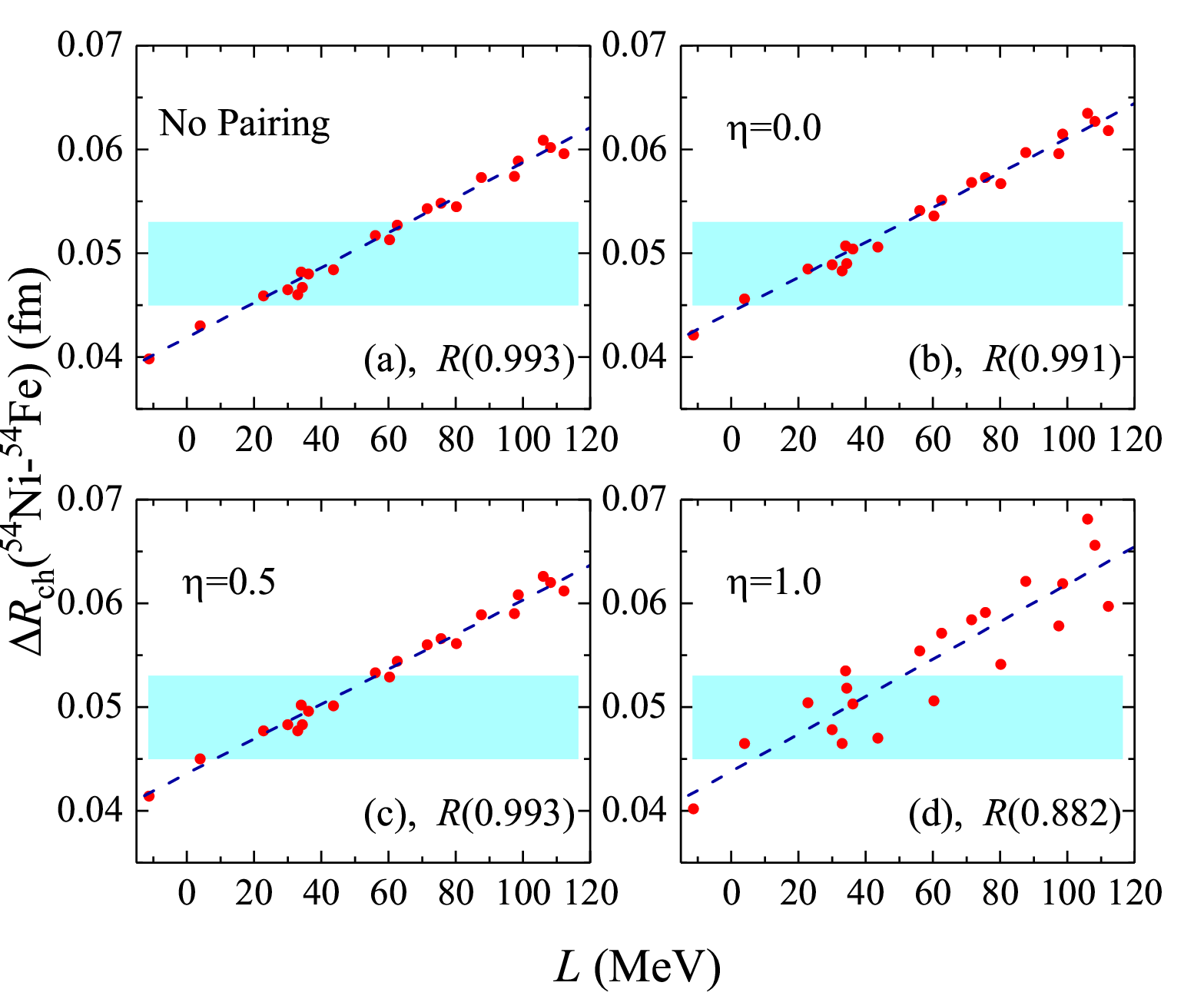}
\caption{(Color online) Charge radii difference $\Delta{R_{\mathrm{ch}}}$ of mirror partner nuclei $^{54}$Ni-$^{54}$Fe as a function of slope parameter $L$ is depicted under various pairing forces: no pairing (a), volume type $\eta=0.0$ (b), mixed type $\eta=0.5$ (c), and surface type $\eta=1.0$ (d) pairing forces. The Pearson coefficients $R$ are presented in parentheses. The database is taken from Refs.~\cite{ANGELI201369,LI2021101440,PhysRevLett.127.182503}, and the systematic error bands are represented in light blue.} \label{fig1}
\end{figure}
In Fig.~\ref{fig1}, plots for the charge radii difference $\Delta{R_{\mathrm{ch}}}$ in $^{54}$Ni-$^{54}$Fe against symmetry energy slope $L$ are shown for various pairing forces.
For the convenience of discussion, the Pearson coefficient $R$, which is used to measure the degree of correlation between two variables, is employed in our study.
Without considering the pairing correlations, the Pearson coefficient $R$ between the $\Delta{R_{\mathrm{ch}}}$ of mirror partner nuclei $^{54}$Ni-$^{54}$Fe and $L$ is determined to be $0.993$. The same scenario can also be found in mixed-type pairing forces. For the volume-type pairing force, the Pearson coefficient $R$ is almost equivalent to those for the no pairing and $\eta=0.0$ cases.
As demonstrated in Ref.~\cite{PhysRevC.107.034319}, the linear correlation between $\Delta{R_{\mathrm{ch}}}$ and $L$ is decreased by the pairing effects.
From Fig.~\ref{fig1}, one can find that the surface-type pairing force actually decreases the linear correlation between $\Delta{R_{\mathrm{ch}}}$ and $L$ in mirror pair nuclei $^{54}$Ni-$^{54}$Fe, where the Pearson coefficient $R$ falls to $0.882$.
Actually, more particles are scattered into the higher single levels under the surface pairing force. This leads to the rather poor correlation between the difference of charge radii of mirror partner nuclei and the slope parameter $L$.

\begin{figure}[htbp]
\includegraphics[scale=0.33]{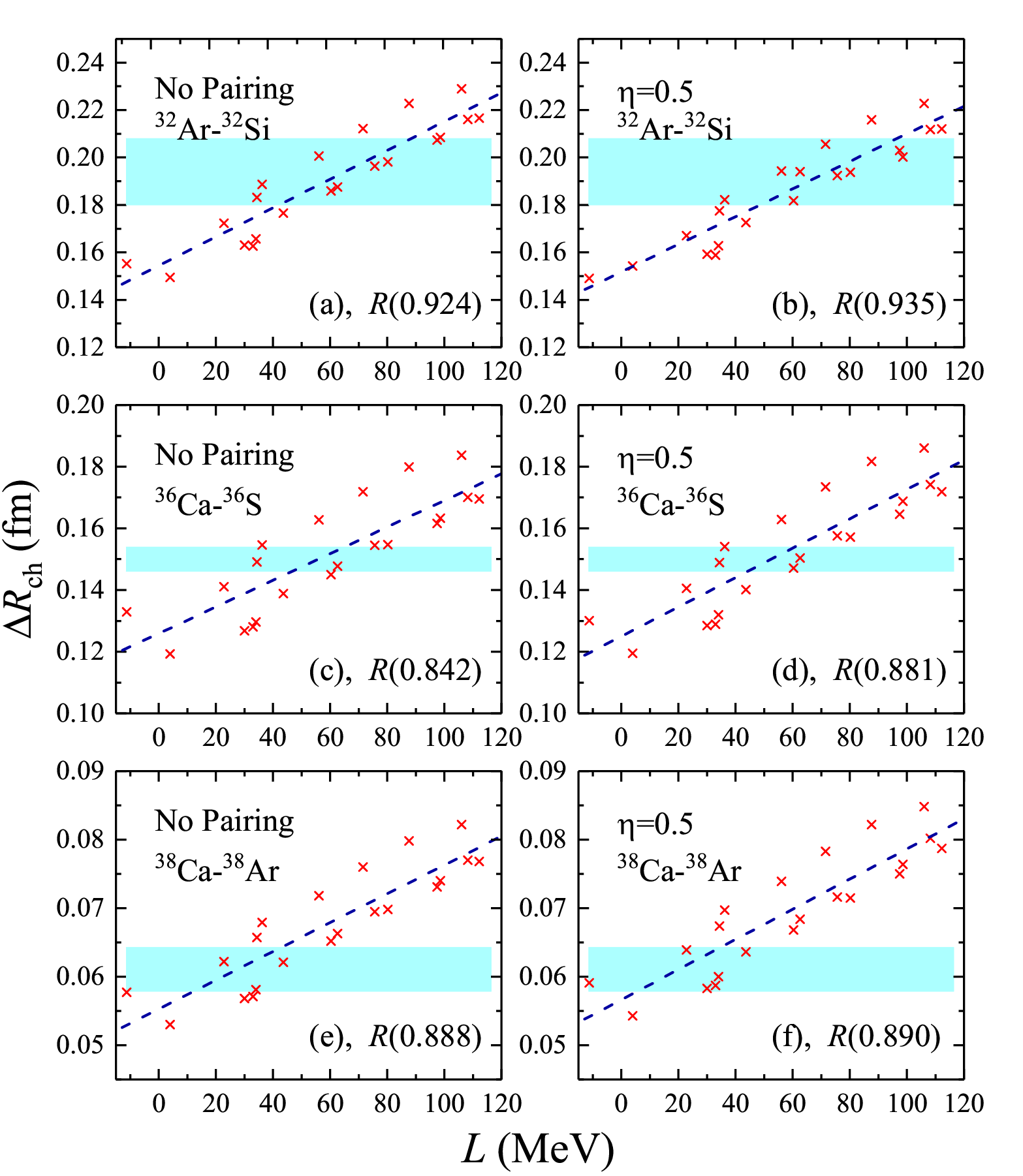}
\caption{(Color online) Charge radii differences $\Delta{R_{\mathrm{ch}}}$ of mirror partner nuclei $^{32}$Ar-$^{32}$Si, $^{36}$Ca-$^{36}$S, and $^{38}$Ca-$^{38}$Ar as a function of slope parameter $L$ depicted under no pairing and mixed-type $\eta=0.5$ pairing forces. The Pearson coefficients $R$ are presented in parentheses. The database is taken from Refs.~\cite{ANGELI201369,LI2021101440,PhysRevResearch.2.022035,Konig:2023rwe}, and the systematic error bands are represented in light blue.} \label{fig2}
\end{figure}
For further focusing on the linear correlation between $\Delta{R_{\mathrm{ch}}}$ and $L$, the charge radii differences $\Delta{R_{\mathrm{ch}}}$ of mirror partner nuclei $^{32}$Ar-$^{32}$Si, $^{36}$Ca-$^{36}$S, and $^{38}$Ca-$^{38}$Ar against slope parameter $L$ are shown in Fig.~\ref{fig2}.
In this figure, clear linear correlations can be observed for these mirror-pair nuclei ($^{32}$Ar-$^{32}$Si, $^{36}$Ca-$^{36}$S, and $^{38}$Ca-$^{38}$Ar).
For the volume-type pairing force, the Pearson coefficients show almost no change in the mirror partner nuclei $^{32}$Ar-$^{32}$Si, $^{36}$Ca-$^{36}$S, and $^{38}$Ca-$^{38}$Ar, with $R$ values of 0.926, 0.881, and 0.836, respectively.
Considering the mixed-type pairing force, the Pearson coefficients $R$ are slightly improved with respect to the no pairing case.
Actually, the Pearson coefficients $R$ are rather poor if the surface pairing force is taken into account.
The corresponding figures are not shown here, but the Pearson coefficients $R$ for the surface-type pairing force are $0.887$, $0.791$, and $0.743$ for the mirror partner nuclei $^{32}$Ar-$^{32}$Si, $^{36}$Ca-$^{36}$S, and $^{38}$Ca-$^{38}$Ar, respectively.
This means that the pairing effect can decrease the linear correlation between $\Delta{R_{\mathrm{ch}}}$ and $L$ if the pairing interactions are only addressed by the surface-type pairing force.

\begin{figure}[htbp]
\includegraphics[scale=0.5]{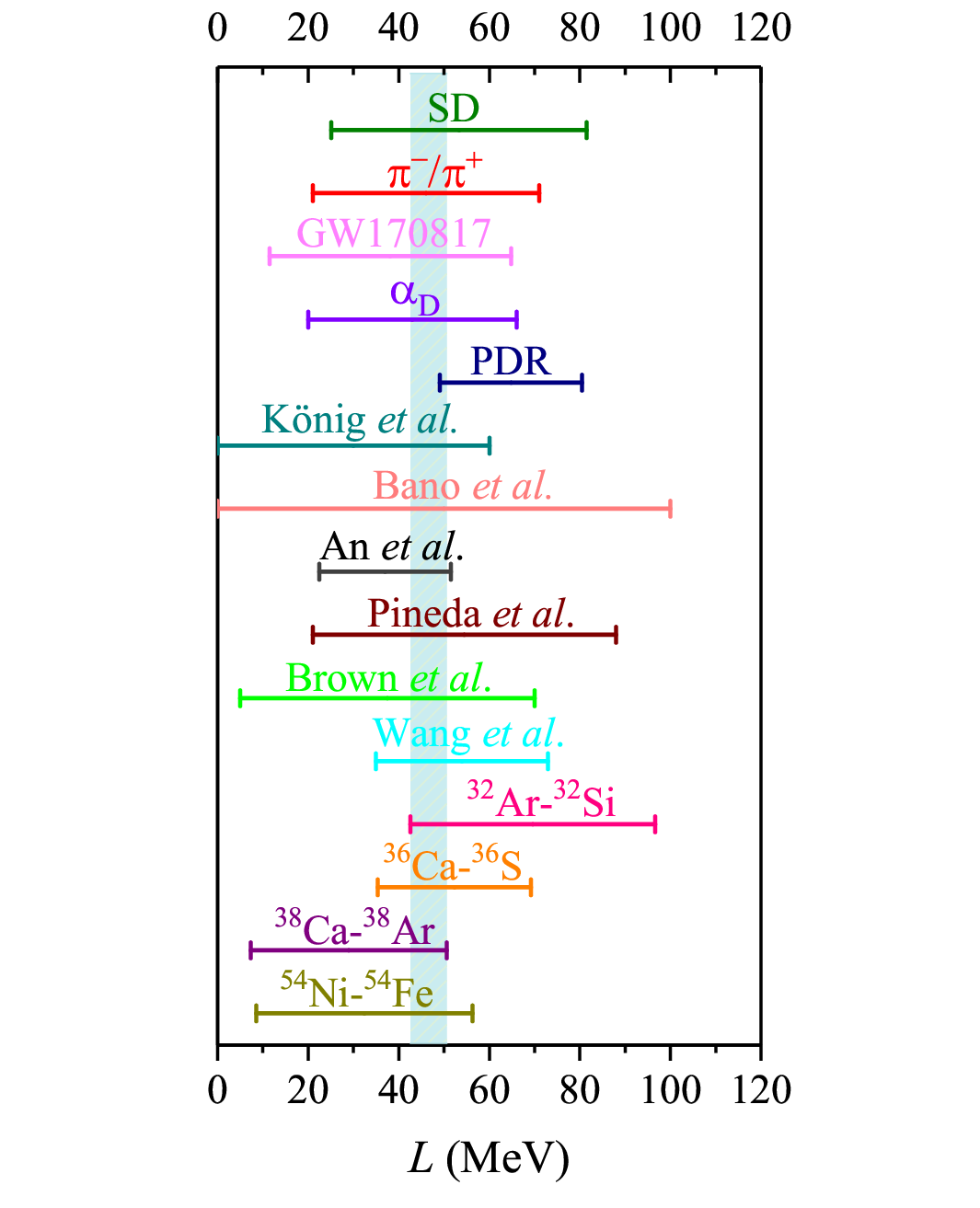}
\caption{(Color online) Plots of the values of slope parameter $L$ obtained in this work ($^{54}$Ni-$^{54}$Fe, $^{38}$Ca-$^{38}$Ar, $^{36}$Ca-$^{36}$S, and $^{32}$Ar-$^{32}$Si) and in previous studies.
The partially enumerated values of $L$ from Wang $et~al$.~\cite{PhysRevC.88.011301}, Brown $et~al$.~\cite{PhysRevResearch.2.022035}, Pineda $et~al$.~\cite{PhysRevLett.127.182503}, An $et~al$.~\cite{nuclscitech34.119}, Bano $et~al$.~\cite{PhysRevC.108.015802}, K\"{o}nig $et~al$.~\cite{Konig:2023rwe}, pygmy dipole resonance (PDR)~\cite{PhysRevC.81.041301},
electric dipole polarizability $\alpha_{\mathrm{D}}$~\cite{PhysRevC.92.064304},  GW170817~\cite{Raithel_2019}, pion production ($\pi^{-}/\pi^{+}$)~\cite{Liu:2023ojj}, and charge exchange spin-dipole (SD) excitations~\cite{Cheng_2023}, are also presented. The color band represents the slope parameter $L=42.57$-$50.64$ MeV obtained in this study. } \label{fig3}
\end{figure}
As mentioned in Refs.~\cite{PhysRevC.88.011301,PhysRevLett.119.122502,PhysRevResearch.2.022035,PhysRevLett.127.182503,Konig:2023rwe}, charge radii differences in mirror partner nuclei provide an alternative approach to pin down the isovector components in the EoS of asymmetric nuclear matter.
To analyze the relatively clear linear correlation between $\Delta{R_{\mathrm{ch}}}$ and $L$, the mixed-type pairing force is employed in the following discussion.
In Fig.~\ref{fig3}, the deduced slope parameters $L$ from the differences of charge radii of mirror-pair nuclei $^{54}$Ni-$^{54}$Fe, $^{38}$Ca-$^{38}$Ar, $^{36}$Ca-$^{36}$S, and $^{32}$Ar-$^{32}$Si are drawn by shaping the range covered in Fig.~\ref{fig2}. Combing the experimental data, the extracted slope parameter falls into the range of $L=42.57$-$50.64$ MeV.
From this figure, one can find that the deduced value of $L$ covers the uncertainty range derived from the mirror charge radii differences~\cite{PhysRevC.88.011301,PhysRevResearch.2.022035,PhysRevLett.127.182503,nuclscitech34.119,PhysRevC.108.015802,Konig:2023rwe}.
Here, the upper limit range of $L$ coming from $^{32}$Ar-$^{32}$Si is 96.65 MeV.
This result is overestimated in comparison with that in Ref.~\cite{Konig:2023rwe}, where the upper range is confined to $L\leq60$ MeV.
This may be due to the fact that shape deformation is ignored in our study.
Besides, the pairing correlations and the limited Skyrme parameter sets used here may also have an influence on evaluating the interval range of slope parameter $L$.

The value of the symmetry energy slope $L$ extracted from the pygmy dipole resonance (PDR) locates at the interval range of $L=64.8\pm15.7$ MeV~\cite{PhysRevC.81.041301}.
The covered range is very narrow compared with that of the other cases depicted in this figure.
Furthermore, the $L$ obtained in this work can cover the uncertainty ranges derived from the nuclear electric dipole polarizability $\alpha_{\mathrm{D}}$~\cite{PhysRevC.92.064304}, dense astrophysical event GW170817~\cite{Raithel_2019}, pion production ratios in heavy-ion collisions~\cite{Liu:2023ojj}, and charge exchange spin-dipole (SD) excitations~\cite{Cheng_2023}.
Besides, this deduced result covers most of the range of $L$ presented in Ref.~\cite{PhysRevC.90.064317}, in which the slope parameter captures the range of $L=47.3\pm7.8$ MeV.
In summary, the extracted range of slope parameter $L=42.57$-$50.64$ MeV means that the rather soft EoS of the isospin asymmetric nuclear matter can be obtained from the implication of charge radii difference in mirror-pair nuclei.

\begin{figure}[htbp]
\includegraphics[scale=0.34]{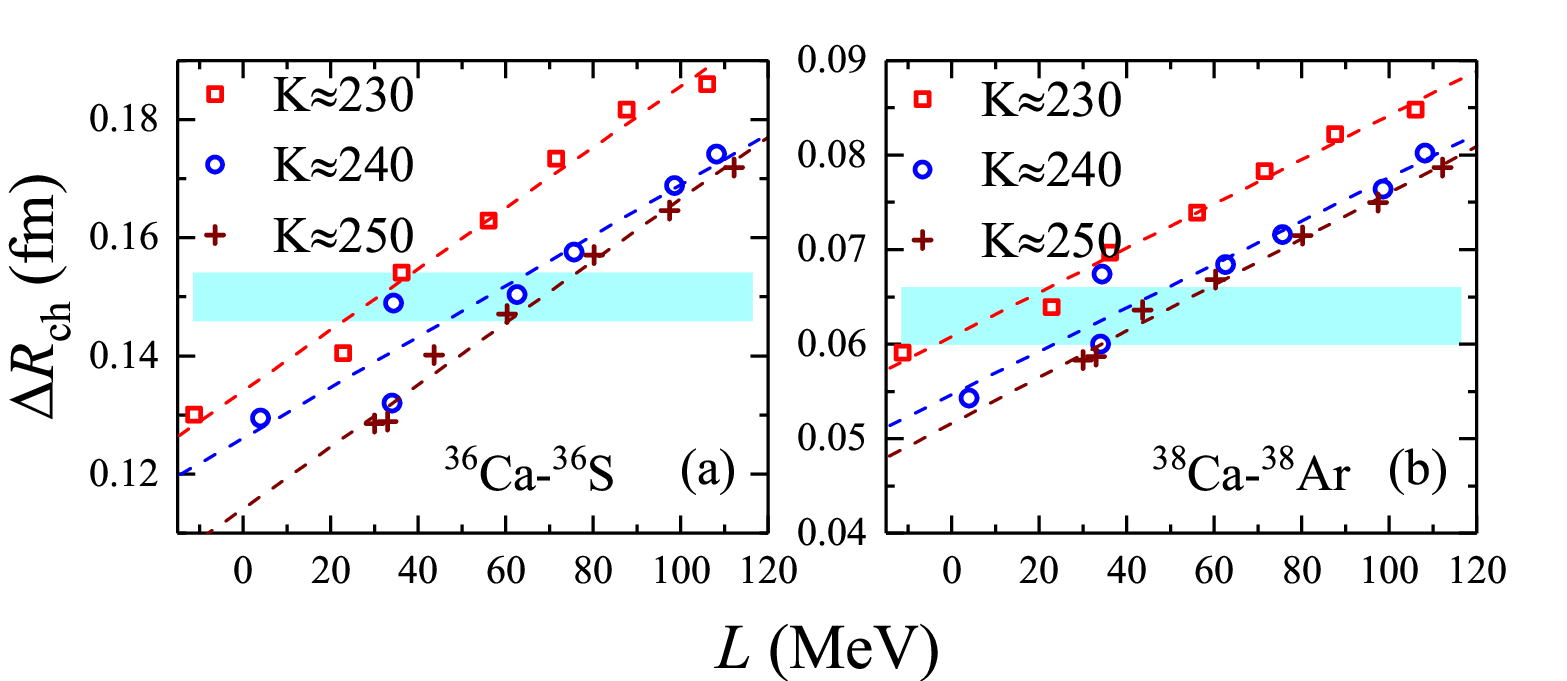}
\caption{(Color online) Charge radii differences $\Delta{R_{\mathrm{ch}}}$ of mirror partner nuclei $^{36}$Ca-$^{36}$S and $^{38}$Ca-$^{38}$Ar plotted as a function of slope parameter $L$ classified by various isoscalar incompressibility coefficients $K$. The corresponding experimental data are taken from Ref.~\cite{PhysRevResearch.2.022035}.} \label{fig4}
\end{figure}
As shown in Fig.~\ref{fig2}, the Pearson coefficients become lower for the mirror-pair nuclei $^{36}$Ca-$^{36}$S and $^{38}$Ca-$^{38}$Ar.
This suggests that the profoundly theoretical uncertainties are encountered in the calibrating protocol.
To clarify this phenomenon, the charge radii differences $\Delta{R_{\mathrm{ch}}}$ of the mirror partner nuclei $^{36}$Ca-$^{36}$S and $^{38}$Ca-$^{38}$Ar against the slope parameter $L$ classified by various isoscalar incompressibility coefficients $K$ are shown in Fig.~\ref{fig4}.
From both panels of Fig.~\ref{fig4}, one can find that the isoscalar incompressibility has an influence on the charge radii difference of mirror partner nuclei. Even from $K\approx230$ to $K\approx250$ MeV, the deviated range for $L$ is approximately $38.0$ MeV.
Here, as shown in Fig.~\ref{fig3}, we mention that the influence of the isoscalar incompressibility coefficient cannot be taken into account in evaluating the slope parameter $L$.
Significantly the isoscalar incompressibility may be indispensable in the calibrating protocol.
As is well known, the accurate description of the density-dependent behavior of symmetry energy plays an important role in recognizing various physical mechanisms through nuclear models, such as the production of superheavy elements~\cite{PhysRevC.94.064608} and cluster radioactivity~\cite{PhysRevC.90.064310,PhysRevC.108.L021303}.
In particular, a recent study suggests that the effective proton-neutron chemical potential difference of neutron-rich nuclei is strongly sensitive to the symmetry energy~\cite{qiu2023bayesian}.
Therefore, more aspects should be taken into account in constraining the slope parameter $L$.

\section{summary and outlook}\label{sec3}
In this work, the correlation between the charge radii differences ($\Delta{R_{\mathrm{ch}}}$) in mirror-pair nuclei $^{32}$Ar-$^{32}$Si, $^{36}$Ca-$^{36}$S, $^{38}$Ca-$^{38}$Ar, and $^{54}$Ni-$^{54}$Fe and slope parameter $L$ at saturation density is investigated based on the spherical Skyrme EDFs. The calculated results suggest that the surface pairing force can further decrease the linear correlation between $\Delta{R_{\mathrm{ch}}}$ and $L$. The deduced slope parameter locates at the interval range of $L=42.57$-$50.64$ MeV under the mixed pairing interaction, that is, the rather EoS. This is in aggrement with most of the previous studies.
Meanwhile, the influence of the isoscalar nuclear matter property on $\Delta{R_{\mathrm{ch}}}$ inevitably constrains the nuclear matter EoS. This may suggest that most of the parametrization forces are ruled out in the calibrated procedure.
As shown in Fig.~\ref{fig4}, the correlation between the charge radii difference of mirror nuclei and the slope parameter of symmetry energy is influenced by nuclear incompressibility.
This seems to suggest that the Skyrme parameter sets adjusted to different slope parameters under a specific incompressibility coefficient can result in a highly linear correlation between the charge radii difference of mirror partner nuclei and the slope parameter $L$. Meanwhile, this may provide an alternative approach to evaluate the possible influential quantities in determining the EoS of asymmetric nuclear matter.
As shown in Ref.~\cite{PhysRevC.107.034319}, the calculated results obtained by density energy functionals consisting of Skyrme and covariant models suggest that more mirror-pair nuclei are ruled out in constraining the slope parameter $L$, but the superior candidates of mirror-pair nuclei $^{44}$Cr-$^{44}$Ca and $^{46}$Fe-$^{46}$Ca may be used to ascertain the limit range of $L$.
Thus, more discussion should be performed in the proceeding work.

Recent studies have shown that the charge-changing cross section measurements can be used to determine the EoS of nuclear matter with mirror nuclei~\cite{XU2022137333}.
This method can also be employed to investigate the charge radii of exotic nuclei~\cite{Wang:2023ugs} and the isospin interactions~\cite{Zhao:2023hcc}.
This provides an accessable approach to ascertain the nuclear symmetry energy.
From the aspect of the charge radii difference in mirror-pair nuclei, the underlying mechanism should be taken into account in determining the nuclear symmetry energy, such as
the local relations among the adjacent neighbouring nuclei~\cite{PhysRevC.90.054318,PhysRevC.94.064315,PhysRevC.102.014306,PhysRevC.104.014303} and $\alpha$-cluster formation~\cite{PhysRevC.89.024318,PhysRevC.87.024310}.
Thus more high-precision charge radii data are urgently required in experimental and theoretical studies.

\begin{acknowledgments}
This work is supported by the National Natural Science Foundation of China under Grants No. 12275025, No. 11975096 and the Fundamental Research Funds for the Central Universities (2020NTST06).
\end{acknowledgments}

\bibliography{refsnew}
\end{document}